\documentclass[letter]{aa}    

\usepackage{graphicx}
\usepackage{txfonts}
\usepackage{lipsum}
\usepackage{subcaption}      
\usepackage{lscape}           
\usepackage{placeins}         

\usepackage{natbib}
\bibpunct{(}{)}{;}{a}{}{,}
                               
\begin{document}

   \title{Discovery of a giant radio outburst of the narrow-line Seyfert 1 galaxy SDSS\,J110546.07$+$145202.4}

\titlerunning{Giant radio outburst in a NLS1 galaxy}
   
   \author{K. \'E. Gab\'anyi\inst{1, 2, 3, 4, 5}\fnmsep\thanks{K\'EG and SK contributed equally to this work}, S. Komossa\inst{6}$^{,\star}$,  A. Kraus\inst{6}, A. Mez\H{o}si\inst{1}, S. Frey\inst{3, 4, 7}
        }

   \institute{Department of Astronomy, Institute of Physics and Astronomy, ELTE E\"otv\"os Lor\'and University, P\'azm\'any P\'eter s\'et\'any 1/A, H-1117 Budapest, Hungary 
            \email{k.gabanyi@astro.elte.hu} \and
             HUN-REN--ELTE Extragalactic Astrophysics Research Group, ELTE E\"otv\"os Lor\'and University, P\'azm\'any P\'eter s\'et\'any 1/A, 1117 Budapest, Hungary 
             \and 
              Konkoly Observatory, HUN-REN Research Centre for Astronomy and Earth Sciences, Konkoly-Thege Mikl\'os \'ut 15-17, 1121 Budapest, Hungary 
              \and
              CSFK, MTA Centre of Excellence, Konkoly-Thege Mikl\'os \'ut 15-17, 1121 Budapest, Hungary
              \and
              Institute of Astronomy, Faculty of Physics, Astronomy and Informatics, Nicolaus Copernicus University, Grudzi\c{a}dzka 5, 87-100 Toru\'n, Poland
\and Max-Planck-Institut für Radioastronomie, Auf dem Hügel 69, D-53121 Bonn, Germany
\and
Institute of Physics and Astronomy, ELTE Eötvös Loránd University, Pázmány Péter sétány 1/A, H-1117 Budapest, Hungary
 }

   \date{Received 7 August 2025}

  \abstract
{}
{We have identified a high-amplitude radio outburst in the course of a large-sample study of the radio properties of narrow-line Seyfert 1 (NLS1) galaxies. }
{We have analysed previous radio data and obtained new radio observations with the Effelsberg $100$\,m telescope, in order to measure the properties and understand the nature of the high-amplitude radio variability. We have also searched for signs of variability in the infrared and optical bands using archival data.}
   {We report the discovery of a rare high-amplitude radio outburst of a NLS1 galaxy,  SDSS\,J110546.07$+$145202.4, with an amplitude of a factor of $>20$ at centimetre wavelengths within $18$\,yr, and remaining at high-state for at least $7.6$\,yr.
Thus, the object transitioned to a radio-loud state with a radio-loudness parameter exceeding $150$. The radio spectrum measured at gigahertz frequencies during the 2020s is flat. We did not find indications of a similar increase in brightness in optical surveys or in the infrared measurements of the {\it Wide-field Infrared Survey Explorer}.}
   {The variability characteristics are inconsistent with tidal disruption events, and hard to reconcile with blazar variability.}


   \maketitle

\section{Introduction}

Radio transients provide us with important insights into the physics of jet formation and evolution, and of particle acceleration under extreme conditions. The class of long-lived radio transients includes stellar tidal disruption events (TDEs), supernovae, gamma-ray burst afterglows, and high-amplitude blazar outbursts, probing different regimes of black hole masses and environments.

Narrow-line Seyfert 1 (NLS1) galaxies are a subgroup of active galactic nuclei (AGNs) with exceptional multi-wavelength (MWL) properties, at one extreme end of the AGN correlation space \citep{Sulentic2000, Grupe2004, Marziani2018}. They are defined by FWHM(H$\beta_\mathrm{broad}$) $< 2000$\,km\,s$^{-1}$, [OIII]/H$\beta < 3$, and strong FeII emission. Main drivers of the NLS1 properties are thought to be high (near-Eddington) accretion onto low-mass supermassive black holes
\citep[SMBHs; review by][]{Komossa2008-rev}. The majority of NLS1 galaxies are radio-quiet. Only $\sim 7$\,\% of them are found to be radio-loud, much fewer than broad-line (BL) AGN \citep{Komossa2006}. 
Note that, formally, the narrow-line type 1 AGN population is composed of Seyfert galaxies (low luminosity) and quasars (high luminosity). For simplicity, we refer to the whole population as NLS1 galaxies. This includes narrow-line type 1 quasars. 

A fraction of radio-loud NLS1 galaxies show radio properties similar to those of blazars, including rapid variability and superluminal motion \citep[e.g.][]{Komossa2006, Yuan2008}. These host relativistic jets.
Their blazar nature was independently confirmed by the Fermi detection of gamma-ray emission from several systems, including, in some cases, gamma-ray 
flaring \citep{Abdo2009-nls1, Foschini2022}.
The amplitudes of variability in the low-frequency 
radio regime were $10$\,\% up to a factor of $\sim 2$ \citep[][]{Yuan2008}.

The radio properties of NLS1 galaxies in general and high-amplitude radio outbursts in particular shed new light on the physics of jet formation and evolution in a new parameter space.
Here, 
we report the discovery of a high-amplitude radio outburst of the nearby NLS1 
galaxy SDSS\,J110546.07$+$145202.4 (hereafter SDSS\,J1105$+$1452) with an amplitude of a factor of $>23$.
We found the outburst in a systematic study of the VLASS radio properties of NLS1 galaxies in general (Gab\'anyi et al. 2025, in prep.), and a search for high-amplitude radio variability in particular. 

SDSS\,J1105$+$1452 is a NLS1 galaxy identified with SDSS
\citep{York2000} at redshift $z=0.12$ with faint [OIII]5007, 
strong FeII emission, and FWHM(H$\beta_\mathrm{broad}$) $=1498$\,km\,s$^{-1}$ \citep{Sun2015}.
It was included in a number of SDSS large-sample studies but has not yet been explored further as a single object.
\cite{paliya2024} estimated  the mass of its black hole as $\sim 10^{6.4} \mathrm{\,M}_\odot$.
They gave a bolometric luminosity of $L_\mathrm{bol} \sim 2 \cdot 10^{44}$\,erg\,s$^{-1}$, and thus an Eddington ratio of $\sim 0.6$. The absolute $B$-band optical magnitude, $-20.7$, is consistent with a Seyfert-type AGN. 

SDSS\,J1105$+$1452 was detected in the Faint Images of the Radio Sky at Twenty-Centimeters \citep[FIRST;][]{first_white} survey at $1.4$\,GHz as a compact source with a flux density of $(1.39 \pm 0.13)$\,mJy in $1999.96$. 
In X-rays it was detected by ROSAT with a count rate of $0.086$\,cts/s, or an X-ray luminosity of $10^{43.7}$\,erg\,s$^{-1}$ according to \citet{Anderson2007}.

This Letter is structured as follows. In Sect. 2 we present previous and new radio observations that demonstrate the outburst and show that the radio emission is point-like. Section 3 gives MWL properties of the galaxy, followed by a discussion and conclusions in Sect. 4. In the following, we assume a $\Lambda$CDM cosmological model with $H_0=70$\,km\,s$^{-1}$\,Mpc$^{-1}$, $\Omega_\mathrm{m}=0.27$, and $\Omega_\Lambda=0.73$. At the distance of SDSS\,J1105$+$1452, $1''$ corresponds to a projected linear size of $2.169$\,kpc \citep{wright}. We use the radio spectral index, $\alpha$ defined as $S\sim \nu^\alpha$, where $S$ is the flux density and $\nu$ is the observing frequency.

\section{Radio observations}

\subsection{Past radio observations}

SDSS\,J1105$+$1452 was observed multiple times in the radio band. 
In addition to the FIRST observation, it was also included in the shallower NRAO VLA Sky Survey \citep[NVSS, ][]{nvss} but remained undetected (Table \ref{table:radioflux}). It was also undetected in the Green Bank $6$\,cm survey \citep{GB6} conducted between November 1986 and October 1987. More recently, SDSS\,J1105$+$1452 was observed several times in the course of the Very Large Array Sky Survey \citep[VLASS, ][]{vlass_lacy} at $3$\,GHz and Rapid Australian Square Kilometre Array Pathfinder (ASKAP) Continuum Survey \citep[RACS, ][]{McConnell2020} at around $1$\,GHz. 

\begin{table*}
\caption{Radio flux density values of SDSS\,J1105$+$1452.}                 
\label{table:radioflux}   
\centering                       
\begin{tabular}{c c c c c c c}   
\hline\hline              
Epoch & Survey & Frequency & Beam size & Flux density & Peak intensity & Reference \\   
 & & (GHz) & (arcsec) & (mJy) &  (mJy\,beam$^{-1}$) & \\ 
 \hline                      
$1987.3$ & GB6 & $4.85$ & $216 \times 204$ & $\lesssim 22$ & -- & \cite{GB6}\\ 
$1994.5$ & NVSS & $1.4$ & $45$ & $\lesssim 2.5$& -- & \cite{nvss} \\ 
   $1999.96$ & FIRST & $1.4$ & $5.4$ & $1.4 \pm 0.1$ & $1.6 \pm 0.3$ & \cite{Helfand_first}\\    
   $2017.96$ & VLASS & $3$ & $2.5$ & $39.0 \pm 0.4$ & $36.9 \pm 0.4$ &  \\ 
   $2020.33$  & RACS-low &  $0.889$  &  $25$ & $32.1 \pm 2.8$ & $32.3 \pm 0.3$ & \cite{racs-low}\\ 
   $2020.63$ & VLASS - SE & $3$ & $2.5$ & $40.4 \pm 0.4$ & $39.4 \pm 0.4$ &  \\
   $2021.01$ & RACS-mid & $1.367$ & $11.2 \times 9.3$ & $32.4 \pm 2.0$ & $31.4 \pm 1.9$ & \cite{racs-mid}\\ 
   $2021.99$ & RACS-high & $1.655$ & $11.9 \times 8.1$ & $42.7 \pm 4.3$ & $41.6 \pm 4.2$ & \cite{racs-high} \\ 
   $2023.04$& VLASS & $3$ & $2.5$ & $43.1 \pm 0.5$ & $42.6 \pm 0.6$ &  \\ 
 \hline
 $2025.59$ & Effelsberg & $4.95$ & $139$ & $32 \pm 2 $ & -- & this work \\
 $2025.59$ & Effelsberg & $6.75$ & $109$ & $35 \pm 2 $ & -- & this work \\
\hline                                  
\end{tabular}
\tablefoot{In the last column, the references for the appropriate catalogues are listed. In the case of the VLASS, the flux densities and peak intensity values obtained from image fitting are given (see text). For the other surveys, we use the published catalogue values. Below the horizontal line, the most recent Effelsberg flux density measurements are listed.}
\end{table*}

The flux densities of SDSS\,J1105$+$1452 published in the RACS-low, the RACS-mid, and the RACS-high catalogues \citep{racs-low, racs-mid, racs-high} at $0.89$\,GHz, $1.37$\,GHz, and $1.66$\,GHz, respectively, are listed in Table\,\ref{table:radioflux}. The flux density measured in RACS-mid is much higher than the FIRST value or the NVSS upper limit. Since the RACS has larger beam size than the FIRST, this difference theoretically could have been caused by a resolution issue. The RACS-mid has a larger beam size than FIRST; however, we show below that the RACS-mid emission is point-like within the resolution, and this fact immediately implies a very high amplitude of variability.
Thus, SDSS\,J1105$+$1452 brightened more than a factor of $23$ during the $\sim 20$\,yr between the FIRST and the RACS-mid observations. 

\begin{figure}[h!]
   \centering
   \includegraphics[width=\hsize, bb=50 0 500 320, clip]{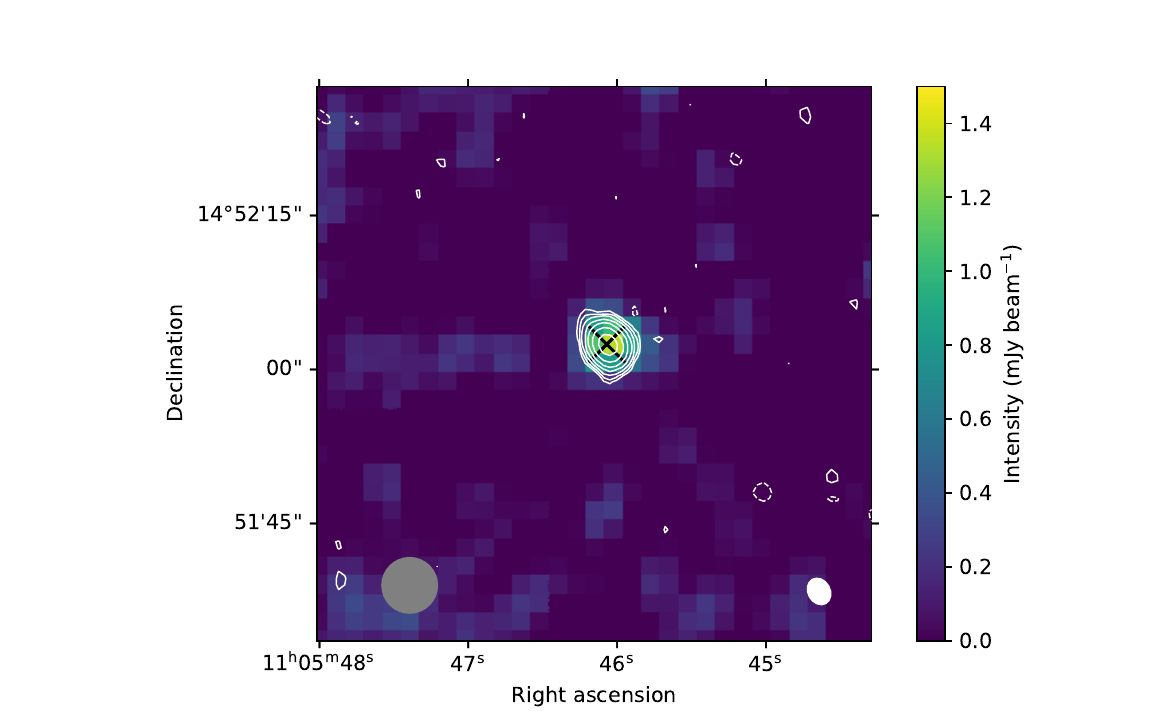}
      \caption{Radio images of SDSS\,J1105$+$1452. The colour scale shows the $1.4$-GHz FIRST image \citep{first_white}. The grey circle shows the restoring beam of the FIRST image in the lower left corner. The white contours represent the $3$-GHz VLASS SE image taken on August 18 2020. The peak intensity is $38.7$\,mJy\,beam$^{-1}$. The lowest contours are drawn at $\pm 0.4$\,mJy\,beam$^{-1}$ (corresponding to a $3 \sigma$ image noise level); further contours increase by a factor of two. The restoring beam of the VLASS SE image is shown as a white ellipse in the lower right corner. Its FWHM size is $2\farcs7 \times 2\farcs2$. The major axis of the position angle is $29\degr$.
      The VLASS and FIRST radio emission is consistent with a point source and no extended jet structure is detected. Further, no other bright radio sources that could have affected the radio measurements are in the surrounding field. The black cross marks the {\it Gaia} DR3 optical position \citep{Gaia_DR3}.}
         \label{fig:radioimage}
   \end{figure}

We have retrieved the VLASS cut-out images of all three available epochs from the Canadian Astronomy Data Centre (CADC\footnote{\url{www.cadc-ccda.hia-iha. nrc-cnrc.gc.ca/en/vlass/}, accessed on 1 Aug. 2025}). For SDSS\,J1105$+$1452, the so-called 'single epoch' (SE) image is available for the second epoch observation (Fig.\,\ref{fig:radioimage}). Since SE images are fully calibrated, and of a higher quality than the quick-look images \citep{vlass_lacy}, we used the SE image instead of the quick-look image for the second epoch. 

We used the National Radio Astronomy Observatory (NRAO) Astronomical Image Processing System \citep[{\sc aips}, ][]{aips} to fit the images with Gaussian components. For all images, the source can be described by a single Gaussian component. The flux densities and peak intensities obtained are listed in Table\,\ref{table:radioflux}. 

\begin{figure}[h!]
   \centering
   \includegraphics[width=\hsize]{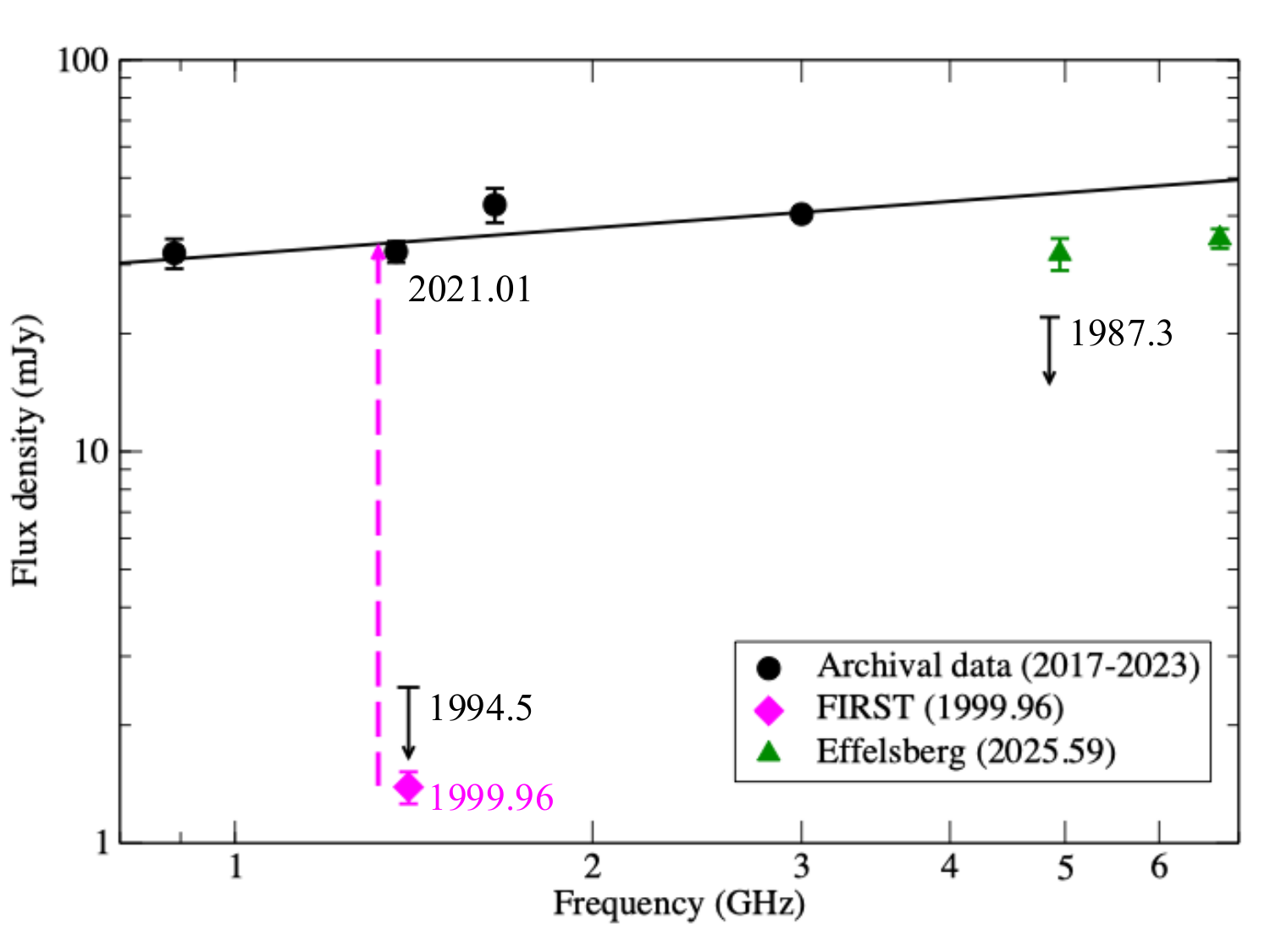}
      \caption{Radio spectrum of SDSS\,J1105$+$1452. The black dots are from the RACS and the VLASS. The solid line is a power law fit to these points. The magenta diamond is from the FIRST. Green triangles show our Effelsberg measurements. The upper limits from NVSS and GB6 are displayed as black downward arrows with observing epochs indicated (for details see, Table\,\ref{table:radioflux}). The dashed magenta line shows the dramatic brightening at $1.4$\,GHz.     
}
         \label{fig:radiospectrum}
   \end{figure}

The three VLASS flux density values match closely, although there is a hint of brightening of a few millijanskys from the first epoch, December 2017, to the last available epoch in January 2023. Nevertheless, to estimate the spectral index we used the average value of the VLASS flux densities, $(40.8 \pm 0.8)$\,mJy. The radio spectrum is flat, and the obtained spectral index is $\alpha = 0.23\pm 0.08$ (Fig.\,\ref{fig:radiospectrum}).

According to the available radio data, SDSS\,J1105$+$1452 is very compact, with a peak-intensity-to-flux-density ratio of $\gtrsim 94$\,\% around $1$\,GHz and at $3$\,GHz observing frequencies as well. The $1.4$-GHz radio power calculated from the RACS-mid observation is $(1.1 \pm 0.1) \cdot 10^{24}$\,W\,Hz$^{-1}$.

\subsection{New Effelsberg radio observations}
To see if the outburst is still ongoing and to measure the radio spectrum beyond $3$\,GHz, we acquired a new radio observation of SDSS\,J1105$+$1452 at the Effelsberg 100\,m telescope on August 4 2025. Following standard data reduction methods \citep{Kraus2003, Komossa2023}, flux densities of $(32 \pm 2)$\,mJy at 4.95 GHz and $(35 \pm 2)$\,mJy at $6.75$ GHz were measured (Table\,\ref{table:radioflux}).

The result shows that the high state is still ongoing and has lasted at least $7.6$\,yr. 
The radio spectrum of SDSS\,J1105$+$1452 does not show a peak until the highest-frequency ($6.75$\,GHz) observation, although we caution that this conclusion is based on non-simultaneous data. The simultaneous Effelsberg flux density measurements give a flat spectrum with $\alpha \sim 0.3$. 

\section{Multi-wavelength properties}

We estimated the radio loudness parameter ($R_{\rm{5 GHz}}$) of SDSS\,J1105$+$1452. $R_{\rm{5 GHz}}$ is defined as the ratio of the $5$-GHz radio flux density to the $4400$\,\AA\,optical flux density \citep{kellermann}. Using the obtained radio spectral index, $\alpha=0.23$, we estimate the $5$-GHz radio flux density to be $45.8$\,mJy, which corresponds to the highest state. The $B$-band optical magnitude of SDSS\,J1105$+$1452 was estimated from the $g$ and $r$ model AB magnitudes from SDSS DR16 \citep{sdssdr16} following the formula used in \cite{paliya2024}. 
The obtained value, $18.12$\,mag, was corrected for Galactic extinction using the NASA/IPAC Galactic Extinction Calculator. For the optical spectral index, we assumed $-0.5$. 
We got a $R_{\rm{5 GHz}}=167$ for the highest state; thus, SDSS\,J1105$+$1452 is a radio-loud object -- at least currently.
The flux ratio of the H$\alpha$ to H$\beta$ emission lines is $\sim 3.4$, implying little dust obscuration in the host galaxy.

We checked multiple photometric data bases to see if the large-amplitude outburst was also recorded in the optical or IR, and to see if we could determine the starting time of the outburst, which must have occurred before 2017.96. SDSS\,J1105$+$1452 was detected by the {\it Wide-field Infrared Survey Explorer} \citep[{\it WISE,} ][]{wise} in all four bands, $3.4\mu$m, $4.6\mu$m, $12\mu$m, and $22\mu$m ($W1$, $W2$, $W3$, $W4$). Its infrared colours, $W1-W2= 0.98$ and $W2-W3=3.25$, place it within the AGN wedge defined by \cite{jarrett2011}.

We downloaded the infrared data measured during the original WISE mission and its continuation, the NEOWISE mission \citep{neowise}, from the NASA/IPAC Infrared Science Archive\footnote{ \url {irsa.ipac.caltech.edu} accessed 2 Aug 2025.} using a search radius of $6''$ around the position of SDSS\,J1105$+$1452. Following the Explanatory Supplement to the WISE All-Sky Data Release Products\footnote{\url{  wise2.ipac.caltech.edu/docs/release/allsky/expsup/index.html}}, we flagged measurement points that have a lower photometric quality, that may be affected by scattered moonlight, or for which the fit indicated extended structure. Finally, we used more than $80$\,\% of the measurements spanning $\sim 14$\,yr. We used the appropriate factors of the given band to convert the magnitudes to flux density values.

No dramatic infrared flux density variability can be seen. SDSS\,J1105$+$1452 experienced a slow decrease of a few percent in its infrared flux densities both in the $3.4\mu$m and $4.6\mu$m bands. At $3.4\mu$m and $4.6\mu$m, it decreased from $\sim 1.9$\,mJy and $2.6$\,mJy measured in 2010 to $\sim 1.7$\,mJy and $2.2$\,mJy measured in 2024.35, respectively. We inspected the individual mission phases each lasting between $1$ to $2$ days for short timescale variability. We did not find significant variability.

In the optical band, we explored the ASAS-SN \citep{Shappee2014} and Catalina sky survey \citep{Catalina2009} archives, with observations since 2014, and 2007, respectively. No systematic rise to any long-lasting, high-amplitude outburst is found.

\section{Discussion and conclusions}

The amplitude of radio variability of SDSS\,J1105$+$1452 at low frequency is exceptionally high, strongly exceeding the amplitudes of NLS1-blazars \citep[e.g.][]{Yuan2008}. At a higher radio frequency of $37$\,GHz, a few NLS1 galaxies were reported to show extreme radio flaring with the Mets\"ahovi radio telescope, but they remained undetected or very faint, $\lesssim 4$\,mJy, at lower
frequencies down to $1$\,GHz with the VLA \citep{Berton2020, Emilia2023}. The cause of the remarkable high-frequency flaring remains uncertain. 
Here, we discuss several possible mechanisms for the low-frequency radio outburst of SDSS\,J1105$+$1452.

TDE: A fraction of TDEs has been detected in the radio regime \citep{Alexander2025}. Jet formation and the actual radio brightness also depend on the presence and density of the circumnuclear interstellar medium (ISM). Radio emission of the known TDEs typically peaks $\sim 1$\,yr after the optical event and then declines, which is very different from SDSS\,J1105$+$1452, which remains radio-bright for at least $7.6$\,yr, suggesting a constant, long-lasting power input. 
In the available optical spectrum, we do not find hints of TDE-like emission lines such as broad HeII. Without any positive evidence of a rare TDE, and given that SDSS\,J1105$+$1452 already was a radio AGN before the outburst, we consider a TDE interpretation unlikely.  

AGN-related variability; star formation estimate at low state:
Assuming that the low-state $1.4$-GHz flux density comes solely from star formation, the FIRST flux density would imply a star formation rate of $26.5$\,M$_\odot$\,yr$^{-1}$ \citep{Hopkins2003}. An upper limit for the star formation rate can also be obtained from the $24\mu$m flux density neglecting the contribution from the AGN and possible jet emission. Using the mid-infrared spectral index from the WISE data, $-1.3\pm0.05$, to estimate the $24\mu$m flux density and then using the formula of \cite{Rieke2009}, the star formation rate is $\lesssim 23$\,M$_\odot$. Thus, the low-state FIRST flux density suggests the existence of emission at AGN luminosities.
 
Blazar variability: The fact that SDSS\,J1105$+$1452 is a radio AGN and of type 1 (near face-on view) suggests that the outburst is related to blazar activity; for instance, in the form of the launch of a major new jet component, or beaming in a jet sweeping our line of sight, or expansion of a compact `young' jet encountering dense ISM. The radio spectrum in the outburst, $\alpha\sim0.2$, is flat, consistent with the core emission of blazars. 
However, the actual long-term light curve pattern is different from the common blazar variability often, but not always, characterised by repeated shorter-lived flaring activity. The short-term radio light curves from year to year still show characteristically smaller amplitude variability; the emission is not constant. 
As highly variable objects across the whole electromagnetic spectrum, blazars often (but not always) show significant flares in the infrared as well \citep{anjum2020}. SDSS\,J1105$+$1452 does not exhibit short timescale variability in any of the $20$ ($\sim 1-2$\,day-long) WISE infrared light curves, or significant flares on longer timescales. Instead, a long-term decrease in the infrared flux density can be seen.

We speculate that a change in the accretion mode could have triggered a longer-lived change in jet activity, explaining the high amplitude of variability (a factor $>23$) and the long duration of the high state, of at least $7.6$\,yr.  Such changing-look events have been frequently observed in the optical spectra of BL type 1 AGNs in recent years, but rarely so far in the optical spectra of NLS1 galaxies \citep{Komossa2024} and not yet in the radio regime of NLS1 galaxies. 

High-resolution radio follow-up observations will constrain the compactness of the jet, and the Rubin Observatory Legacy Survey of Space and Time will provide a well-covered optical light curve in which to search for optical variability associated with future high-amplitude radio variability. Many more radio transients are expected to be discovered in upcoming radio surveys with the Square Kilometre Array. High-amplitude radio outbursts in NLS1 galaxies such as the one we have reported for SDSS\,J1105$+$1452  and their rapid MWL follow-ups will then provide us with important new insights into the physics of the central engine and the disc--jet system in particular.

\begin{acknowledgements}
    This work is partly based on data obtained with the 100\,m telescope of the Max-Planck-Institut f\"ur Radioastronomie at Effelsberg. This scientific work uses data obtained from Inyarrimanha Ilgari Bundara, the CSIRO Murchison Radio-astronomy Observatory. We acknowledge the Wajarri Yamaji People as the Traditional Owners and native title holders of the Observatory site. CSIRO’s ASKAP radio telescope is part of the Australia Telescope National Facility (\url {https://ror.org/05qajvd42}). Operation of ASKAP is funded by the Australian Government with support from the National Collaborative Research Infrastructure Strategy. ASKAP uses the resources of the Pawsey Supercomputing Research Centre. Establishment of ASKAP, Inyarrimanha Ilgari Bundara, the CSIRO Murchison Radio-astronomy Observatory and the Pawsey Supercomputing Research Centre are initiatives of the Australian Government, with support from the Government of Western Australia and the Science and Industry Endowment Fund.
     This paper includes archived data obtained through the CSIRO ASKAP Science Data Archive, CASDA (\url {http://data.csiro.au}). This research was also supported by HUN-REN, the NKFIH OTKA K134213 grant, and the NKFIH excellence grant TKP2021-NKTA-64.     
\end{acknowledgements}

\bibliographystyle{aa}
\bibliography{ref}

\end{document}